\documentclass[prl,aps,twocolumn,superscriptaddress]{revtex4}

\usepackage{amsbsy,amssymb,amsmath,bm}
\usepackage{epsfig,rotate}
\usepackage{fancyhdr}

\usepackage{graphicx,color}
\usepackage{dcolumn}
\usepackage{bm}
\usepackage{amssymb}

\begin{document}


\title{Degeneracy and  Criticality from Emergent Frustration in Artificial Spin Ice}

\author{Gia-Wei Chern$^1$, Muir J. Morrison$^1$, and Cristiano Nisoli}
\affiliation{{Theoretical Division and Center for Nonlinear Studies, LANL, Los Alamos, NM 87545, USA}}

\date{\today}

\begin{abstract}
Although initially introduced to mimic the spin-ice pyrochlores, no artificial spin ice has yet exhibited the 
expected degenerate ice-phase with critical correlations similar to the celebrated Coulomb phase in the pyrochlore lattice. Here we study a novel artificial spin ice based on a vertex-frustrated rather than pairwise frustrated geometry and show that it exhibits a quasi-critical ice-phase of extensive residual entropy  and, significantly,  algebraic correlations.
Interesting in its own regard as a novel  realization of frustration in a vertex system, our lattice  opens new pathways  to study defects in a critical manifold and to design degeneracy in artificial magnetic nanoarrays, a task so far elusive.
\end{abstract}

\maketitle

Artificial Spin Ice (ASI) has raised considerable interest  for its technological potentials, 
and as a tailorable medium to investigate  collective phenomena in a materials-by-design approach 
\cite{nisoli13,wang06,bramwell06,nisoli07,nisoli10,morgan11, morgan13, moller09,chern11,lammert10,ke08,mengotti09,libal06,tanaka06, qi08,li10,morrison12, kapaklis12,nisoli12,greaves12,farhan13,chern13}.
It was  inspired by the so-called spin ice compounds~\cite{harris91,bramwell01}, a special class of pyrochlore ferromagnets 
 which,  as water ice~\cite{pauling35,petrenko,bernal33}, retain a finite entropy density even at very low temperatures. 
The nontrivial local ordering dictated by the so-called two-in-two-out ``ice rules''~\cite{petrenko} in pyrochlore lattice gives rise to 
dipolar-like power-law spin correlations~\cite{henley10} at large distances. 
A recent surprise is the realization that dipolar excitations in the ice manifold fractionalize into emergent magnetic monopole quasiparticles~\cite{castelnovo08}. Degeneracy is  essential  for magnetic monopoles to play a significant role in (artificial) spin ice. 

The original artificial spin ice presented by Wang {\em et al.} consists of magnetically interacting elongated permalloy nanoislands arranged 
as links of a square lattice~\cite{wang06,nisoli07, nisoli10}. At low temperatures, magnetic configurations satisfying the 2-in-2-out ice rules
in square lattice can be mapped
to a six-vertex model. However the anisotropic nature of magnetic interactions in a 2D arrangement lifts the degeneracy of the six
distinct 2-in-2-out ice-rule vertices. The ground state of square ASI is thus the ordered phase of the $F$-model~\cite{baxter, lieb67}. Proposals to circumvent  this limit~\cite{moller06}, however, present technical challenges in nanofabrication.  
So far, only kagome ASI, with  islands  arranged along the edges of a honeycomb lattice~\cite{tanaka06,qi08,li10}, 
exhibits an extensive degeneracy at the vertex-level description, resulting from the 2-in-1-out or 1-in-2-out pseudo-ice rules~\cite{disclaimer}.
Yet this  pseudo-ice regime is  non-critical, with exponentially-decaying spin 
correlation.

\begin{figure}[b]
\includegraphics[width=0.9\columnwidth]{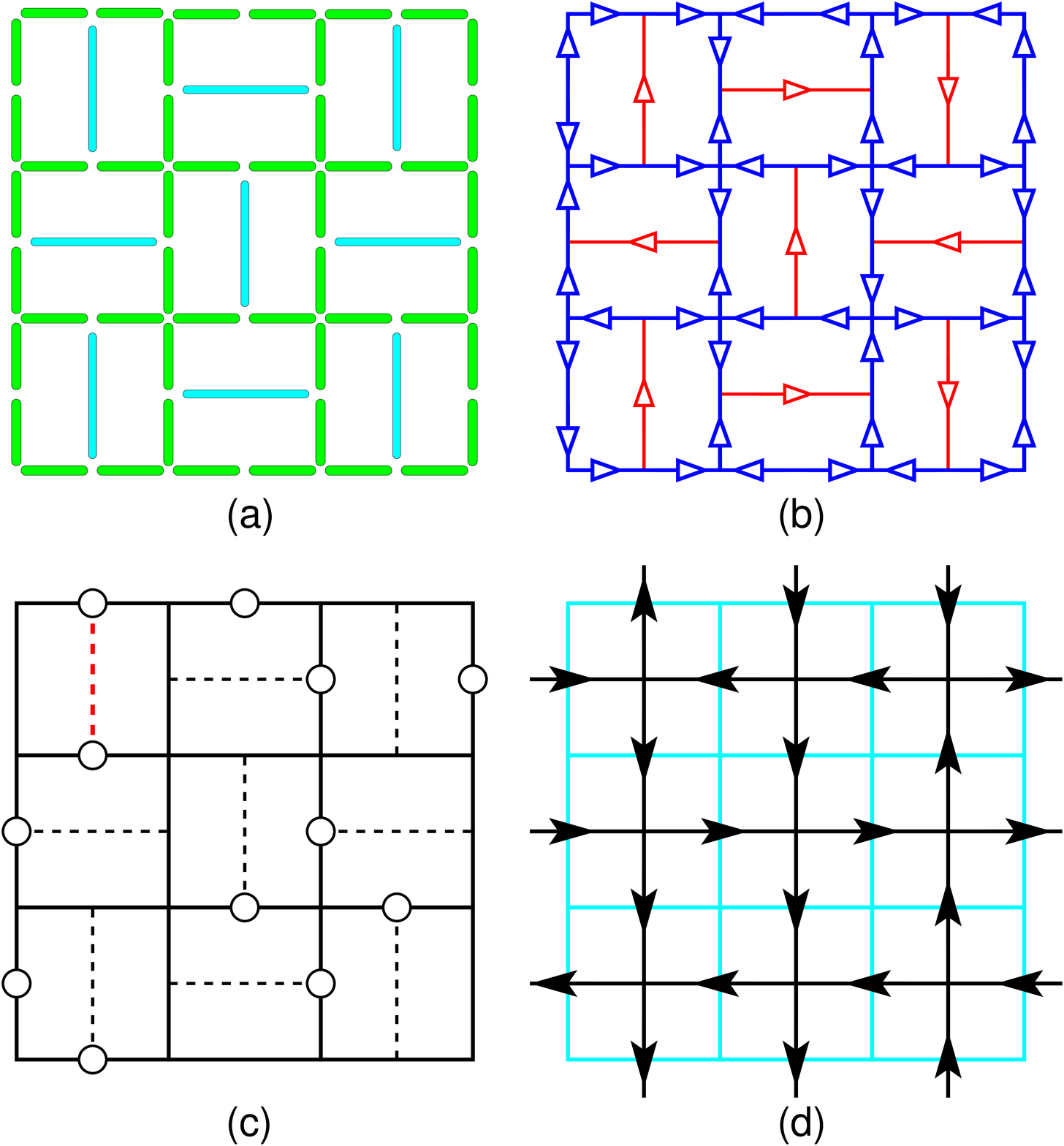}
\caption{(Color online) (a) The lattice geometry of the shakti spin ice. The spin-ice state and the corresponding
defect-vertex configuration in a typical disordered ground state are shown in panels (b) and (c), respectively.
The defect type-II' vertices are indicated by circles. The defect configuration is further mapped to an emergent 6-vertex model in (d).
\label{fig:lattice}}
\end{figure}

In this paper, we show that a critical degenerate phase in ASI, reminiscent of 3D natural spin ice, can be realized by exploiting the concept of an emergent vertex-frustration~\cite{morrison12}, instead of frustrated pairwise interactions, thus solving a long standing problem in the field. 
Specifically, we consider a ``shakti'' lattice shown in Fig.~\ref{fig:lattice}(a) which (as a graph) is 
isomorphic
to the so-called Cairo pentagonal tiling~\cite{cairo}.
The degeneracy of this critical ice manifold follows from the inability of allocating all the vertices in their lowest energy configuration.
This frustration in vertex-allocation is in contrast to the conventional frustrated magnets in which the extensive degeneracy originates from a degeneracy built-in the constituting units, e.g. 
tetrahedra or triangles. Here instead the elementary units, the vertices, are locally ordered, with a unique lowest energy configuration. However an emergent composite unit, the plaquette is frustrated toward the optimal  allocation of all its vertices. We show that the degeneracy of these plaquettes  can be mapped to an exactly solvable {\it thermal} state of an emergent, frustrated six-vertex system, the $F$-model. This emergent phase is known to be critical~\cite{lieb67}.

\begin{figure}[t]
\includegraphics[width=0.99\columnwidth]{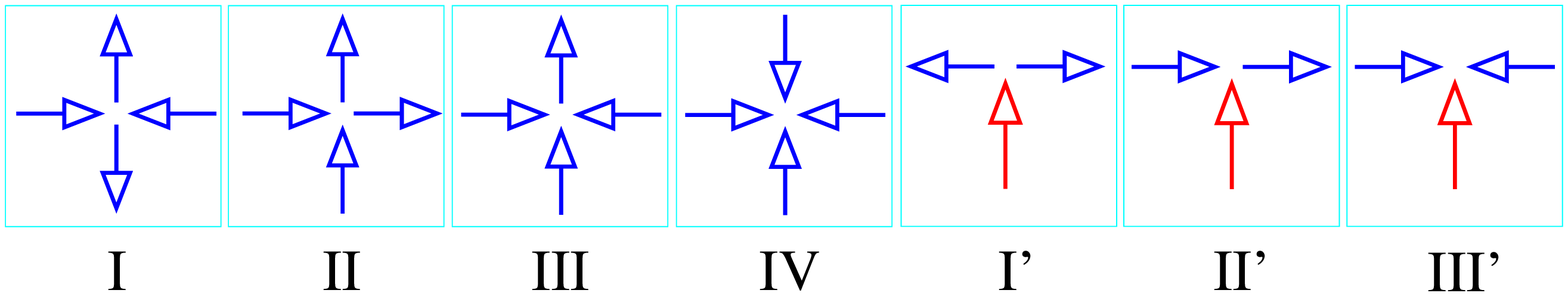}
\caption{(Color online) The seven different types of vertices in artificial spin ice shown in Fig.~\ref{fig:lattice}(a, b), arranged by increasing energy left to right.
The multiplicities of the various vertex types are: $q_{\rm I} = 2$, $q_{\rm II} = 4$, $q_{\rm III} = 8$, and $q_{\rm IV} = 2$
for $z=4$ vertices, and $q_{\rm I'} = 2$, $q_{\rm II'} = 4$, and $q_{\rm III'} = 2$ for $z=3$ vertices.
\label{fig:vertex}}
\end{figure}

The shakti lattice shown in Fig.~\ref{fig:lattice}(a) can be derived from the square lattice~\cite{wang06} by alternatively placing an additional vertical 
or horizontal permalloy island in each square plaquette. The lattice contains vertices with coordination numbers $z=4$ or 3. 
Although the perpendicular geometry of the islands makes different vertex configurations nonequivalent (see Fig.~\ref{fig:vertex}),
an extensive degeneracy is regained as not all vertices can assume their minimum-energy configurations simultaneously.

Since new generation annealing protocols~\cite{morgan11,kapaklis12,nisoli12,greaves12, sheng13, nisoli13} might open a pathway to lower entropy states than the one described by vertex models, 
here we provide a comprehensive numerical study including the full long-range dipolar interactions of the shakti lattice. 
Yet, to understand the origin of the extensive degeneracy, we first focus on the magnetostatic energies of the vertices at the nearest-neighbor level.

Energetically, there are four distinct vertex types for the $z=4$ sites, labeled by numerals I--IV, 
while the three types of $z=3$ vertices are labeled as type I$^\prime$, II$^\prime$, and III$^\prime$ (see Fig.~\ref{fig:vertex}). 
For a realistic implementation, we are interested in energy hierarchies satisfying $\epsilon_{\rm I}<\epsilon_{\rm II}<\epsilon_{\rm III}$, $\epsilon_{\rm I'}<\epsilon_{\rm II'}<\epsilon_{\rm III'}$, and $\epsilon_{\rm II'}-\epsilon_{\rm I'}<\epsilon_{\rm II}-\epsilon_{\rm I}$,
which have been demonstrated experimentally~\cite{nisoli13, wang06, nisoli07, nisoli10,li10}, and which we have further corroborated with micromagnetics simulations (supplementary materials). 
More importantly, the first two conditions therefore ensure there is {\it no degeneracy at the vertex level:} 
the finite residual entropy density of the ice phase (which is the  ground state of the vertex-model) is a consequence of vertex-frustration, 
i.e. the inability of all vertices to reach the lowest-energy states simultaneously. The third condition ensures that the vertex-frustration is accommodated with Type~II$^\prime$, rather than Type~II, vertices. 
We stress that our results, as often in vertex models, only depend on the energy hierarchy of vertices, while the   
particular parametrization is largely irrelevant---a rather significant point of view for practical implementations. 

\begin{figure}[t]
\includegraphics[width=.98\columnwidth]{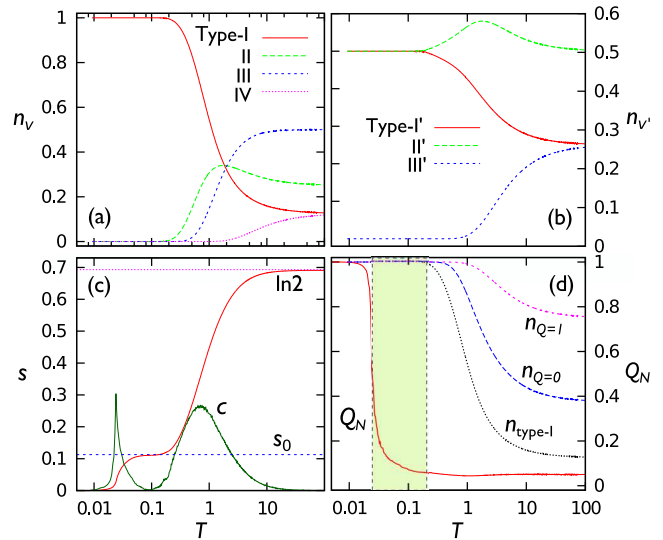}
\caption{(Color online) 
\label{fig:mc} Monte Carlo simulation of the shakti spin ice with $N_s = 1280$ spins
using the dumbbell representation for the computation of magnetostatic energies.
(a) and (b) show the fraction of various vertex types as a function of temperature $T$ measured in units 
of the parameter $\alpha$ from the dumbbell representation. (c) shows the temperature dependence of entropy density $s$ and specific-heat $c$. 
The entropy curve $s(T)$ is obtained by integrating the specific-heat $c(T)/T$. (d) shows the temperature dependence of order parameter $Q_N$ 
characterizing the N\'eel-type charge-order in the ground state; also shown are vertex populations with total charge $Q=0$ and $Q=\pm1$.
The quasi-critical $F$-model phase is indicated by the shaded regime.
}
\end{figure}

To explore potential low-$T$ thermodynamic phases, we perform Monte Carlo simulations using the dumbbell model~\cite{moller09}
which takes into account the full long-range interactions between the nanoislands.
We use the following geometrical parameters: $l_s = 0.475 a$ and $l_c = 0.95 a$, where $l_s$ and $l_c$ are the length of the short
and center nanoislands, respectively, and $a$ is the length of the square plaquette. The dipole moments of the two types of islands are $\mu_c = 3\mu_s$,
and the charges of the dumbbell is given by $Q = \mu/l$.

The simulation results are summarized in Fig.~\ref{fig:mc}.   
At high temperatures the system is in an uncorrelated paramagnetic phase as the populations of various vertex types reach  their respective multiplicities.
A broad peak in the specific-heat [Fig.~\ref{fig:mc}(c)] signals the crossover into an ice phase dominated by vertices of types I, I$^\prime$ and II$^\prime$.
We will show in the following that this ice regime is a critical phase and is described by an exactly solvable $F$-model.
As temperature further decreases, the system undergoes a continuous transition into a phase characterized by a staggered ordering of magnetic charges;
the corresponding order parameter $Q_N$ versus temperature is shown in Fig.~\ref{fig:mc}(d). Details of this long-range ordered state are presented below.
The entropy density $s$ of the system as a function of temperature is obtained by integrating the $c/T$ curve. At high temperatures, the entropy density 
approaches $k_B \ln 2$ as expected for an Ising magnet [Fig.~\ref{fig:mc}(c)]. 
More importantly, a plateau $s_0 \approx 0.1178\,k_B$ appears in the $F$-model regime, indicating an extensive degeneracy of the ice phase.

\begin{figure}
\includegraphics[width=0.95\columnwidth]{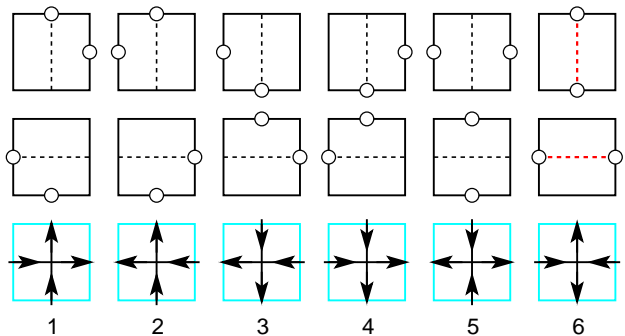}
\caption{(Color online) The six possible two-defects configurations in a plaquette (for both type-A and B)
and their mapping to the 2-in-2-out vertices.
\label{fig:mapping}}
\end{figure}

We first consider the ice phase above the charge-ordering transition, because it is the most likely to be observed experimentally and can be described by vertex considerations. While all $z=4$ sites are minimum-energy type-I vertices in the low-$T$ ice regime, 
only half of the $z=3$ vertices are in the ground state, the other being type-II$^\prime$; see Figs.~\ref{fig:mc}(a) and (b). 
We can thus characterize the spin-ice state via the locations of these unhappy type-II$^\prime$ vertices (or ``defects'') on plaquettes: 
Figs.~\ref{fig:lattice}(b) and (c) show a generic disordered spin-ice state
and the corresponding defect configuration, respectively. This mapping from spins into defects on plaquettes is 
at least 2-to-1: each spin-ice state and its time-reversal partner are mapped to the same defect configuration. Moreover,  when  both defects sit at the two ends of the center spin, i.e. in type-6 plaquette in Fig.~\ref{fig:mapping}, 
there is an additional $Z_2$ degree of freedom associated with the direction of that  (red) spin.

By enumerating all possible magnetic configurations in a plaquette one notes that
each plaquette has at least two defects in any spin-ice state. As a result, any configuration in which each plaquette has {\em exactly} two defects is a ground state,
consistent with the numerical result that $n_{\rm II'} = 1/2$ in the ice phase.
This two-defects constraint is similar to the 2-in-2-out ice rule in square ice. Fig.~\ref{fig:mapping} shows the six different two-defects configurations in a plaquette. 
Since there are two types of plaquettes in our lattice with different orientations of the center island, we refer to plaquettes with vertical (horizontal) center spin  
as type-A (B) plaquettes.
We can bijectively map each defect configuration to an emergent 6-vertex state in the square lattice, by drawing an arrow from a type-A to its neighboring type-B plaquette if their common edge contains a defect vertex; conversely, an edge without an unhappy vertex corresponds to an arrow 
pointing from a type-B to a type-A plaquette [Fig.~\ref{fig:mapping} and also
 Fig.~\ref{fig:lattice}(d)].

The partition function of the emergent 6-vertex model can now be computed  by taking into account the additional doublet degeneracy of type-6 vertices. Let $n_6$ denote the number of type-6 vertices in a given 6-vertex 
configuration $\mathcal{C}$. The ground-state partition function of the shakti spin ice is simply
\begin{equation}
	\label{eq:z1}
	Z = \sum_{\mathcal{C}} 2^{n_6} = \sum_{\mathcal{C}} \bigl(\sqrt{2}\bigr)^{n_5 + n_6}.
\end{equation}
Here we assume periodic boundary conditions. Since type-5 and 6 vertices are sources and sinks, respectively, of the horizontal arrows or `fluxes', 
and the total flux is conserved, thus $n_5 = n_6$~\cite{baxter}. Even though $n_5 = n_6$ is strictly true only for periodic boundary conditions, 
we can expect $n_5 \sim n_6$ for any boundary conditions in the thermodynamic limit.

The degeneracy of Eq.~(\ref{eq:z1}) can be estimated as: $W = W_{\rm ice} \times 2^{N_{\square} \langle n_6 \rangle}$, where
$W_{\rm ice}=\bigl(\frac{4}{3}\bigr)^{3N_{\square}/10}$ is the degeneracy of the ideal square ice~\cite{baxter},
$\langle n_6 \rangle \sim 0.189$ is the average population of type-6 vertex in a ice state, and $N_{\square}$
is the number of plaquettes (or vertices in a 6-vertex state). Since $N_\square = N_s/5$ in shakti ice, 
where $N_s$ is the number of spins, this gives a residual entropy  $s_0 = k_B \ln W/N_s \sim 0.1125 k_B$,  close to 
the numerical result.

The entropy density of the shakti spin-ice can be exactly computed 
by recasting the partition function~(\ref{eq:z1}) into that of the standard $F$-model~\cite{lieb67},
\begin{eqnarray}
	\label{eq:z2}
	Z = \sqrt{2}^{N_\square} \, Z_F = \sqrt{2}^{N_\square}\, \sum_{\mathcal{C}} \prod_{i=1}^6 \omega_i^{n_i},
\end{eqnarray}
where $\omega_i$ denotes the statistical weight of the type-$i$ vertex: $\omega_1 = \omega_2 = \omega_3 = \omega_4 = e^{-K}$ 
and $\omega_5 = \omega_6 = 1$. The case of shakti ice then corresponds to $K = \frac{1}{2} \ln 2$. 
Note that without the $Z_2$ degree of
freedom associated with the center spin in type-6 plaquettes, which could be frozen by application of a suitably small magnetic field, the system becomes Lieb's equal-weight six-vertex ($K = 0$) or square-ice model~\cite{lieb67}. 

By varying the effective temperature $T_{\rm eff} = 1/K$, the $F$-model undergoes a
a Kosterlitz-Thouless transition at $K_c = \ln 2$~\cite{baxter}.
The correlation length is finite in the ordered state for $K > K_c$, while the system remains critical
with an infinite correlation length for $K \le K_c$. Our case with $K = K_c/2$ thus corresponds
to a critical phase with disordered vertices and spins, as anticipated above.

The residual entropy of Eq.~(\ref{eq:z2}) is $S/k_B  = \ln Z = (N_\square/2)\ln 2 +\ln Z_F$.
Computing the entropy density of the $F$-model with $K = K_c/2$ via the Bethe {\em ansatz}~\cite{lieb67}, we have the residual entropy density for the shakti ice,
\begin{eqnarray}
	\label{eq:S}
	 \frac{S}{N_\square\, k_B} = \frac{1}{\pi}\int_0^{\frac{\pi}{2}} \ln\cot(k/2)\,dk = \frac{2G}{\pi} = 0.583122,
\end{eqnarray}
where $G$ is Catalan's constant~\cite{lieb67}. Since the number of spins $N_s = 5 N_{\square}$ in shakti ice, 
Eq.~(\ref{eq:S}) corresponds to an entropy  $S/N_s\,k_B = 0.116624$ per spin,
which is fairly close to our numerical value $s_0 = 0.1178 k_B$ [Fig.~\ref{fig:mc}(c)] as well as to our estimate above.

\begin{figure}
\includegraphics[width=0.98\columnwidth]{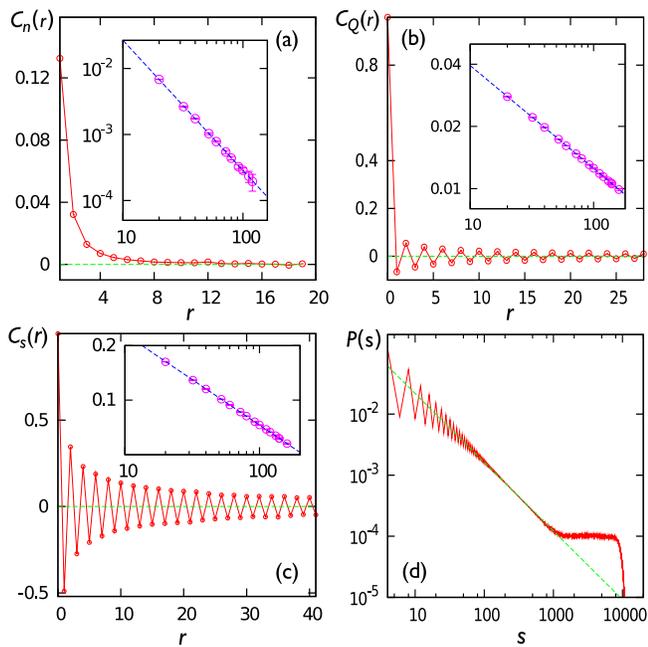}
\caption{(Color online) 
\label{fig:corr} The correlation function for (a) defect-vertex, (b) magnetic charge,
and (c) magnetic moment measured  at the vertical edges of two plaquettes separated by $r$ unit cells along the diagonal direction of the lattice (Monte Carlo simulations using the loop algorithm~\cite{evertz93}). Insets show the same data (with absolute values) in log-log plot, indicating power-law correlations $C(r) \sim r^{-\eta}$ with exponents $\eta_n = 1.96986 \pm 0.0046$, 
$\eta_Q = 0.497144 \pm 0.001952$, and $\eta_s = 0.502075 \pm 0.002028$, respectively.
Panel~(d) shows the probability distribution of loop lengths $s$.}
\end{figure}

As mentioned above, the isomorphism with the $F$-model  implies algebraic correlations,  
which we demonstrate by Monte Carlo simulations (Fig.~\ref{fig:corr}). We employ the efficient loop algorithm in which detailed balance
is always satisfied locally when constructing loops~\cite{evertz93,syljuasen04}.
The defect-defect correlation function is defined as $C_n(\mathbf r) = \langle \delta n(\mathbf r)\,\delta n(0) \rangle$,
where $\delta n(\mathbf r) = n(\mathbf r) - \langle n \rangle$, and $n(\mathbf r) = 1$ or 0 in presence
or absence of defect  at edge $\mathbf r$ and $\langle n \rangle = 1/2$.  Fig.~\ref{fig:corr}(a) shows that  $C_n(r)$ falls off quickly beyond 
a few lattice constants. A stronger correlation  $C_Q(\mathbf r) = \langle Q(\mathbf r)\, Q(0)\rangle$ 
is obtained for the uncompensated magnetic charges $Q = \pm 1$ at the $z=3$ vertices [see Fig.~\ref{fig:corr}(b)].
By reconstructing spin configuration from the defect configuration generated via the loop algorithm, 
we also computed the spin-spin correlation $C_s(\mathbf r)$ shown  in Fig.~\ref{fig:corr}(c). 
In all three cases, the log-log plot demonstrate  algebraic decay.
The oscillations indicate the preferred short-range local order of the ice-phase, as intuitively, a staggered arrangement of magnetic charges should minimize their Coulomb repulsion. We emphasize, however, that this is distinct from the full charge-ordered phase discussed below. 
The lack of a length scale in a critical phase also implies that loops of all lengths will be found in a typical disordered state. 
Indeed, the probability distribution function $P(s)$ of loop length $s$ shown in Fig.~\ref{fig:corr}(d) exhibits a power-law distribution $P(s) \sim s^{-1.125}$ 
for short loops.  A flat distribution at large $s$ is due to winding loops in a finite system~\cite{jaubert11}.


\begin{figure}
\includegraphics[width=0.85\columnwidth]{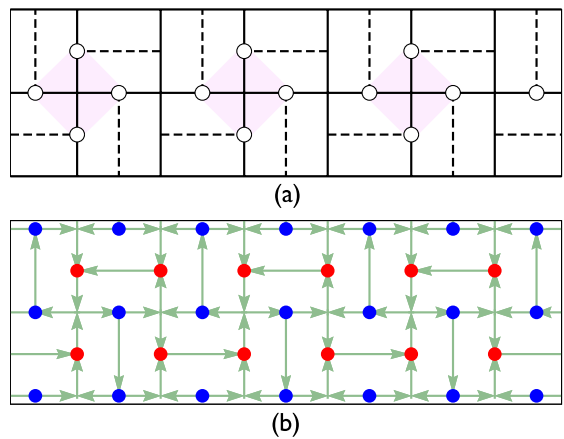}
\caption{(Color online) 
\label{fig:smectic} (a) Clustering of defect vertices along $x$-direction in the charge ordered state. (b) Staggered ordering
of magentic charges and the spin-configuration correspond to the defects pattern in (a). Here blue and red circles
denote $\pm 1$ magnetic charges in natural units.}
\end{figure}

Finally, we discuss the magnetic structure below the charge-ordering transtition~\cite{supple}.
Interestingly, we find that a residual non-extensive degeneracy remains in shakti ice with N\'eel-type charge order;
this residual degeneracy is related to an emergent sliding symmetry~\cite{nussinov}.
In this phase, the defect-vertex configuration exhibits a layered structure in which a period-2 1D-ordering spontaneously appears 
in either $x$ or $y$ directions. The ordering thus reduces the $C_4$ rotation symmetry to $C_2$. 
Fig.~\ref{fig:smectic} shows the crystallization of the defect-vertices along the $x$ direction
and the corresponding spin and charge configuration.
The system remains ``almost degenerate'' energetically as the 1D structure on a given chain is uniformly shifted by one lattice constant~\cite{supple};
the residual degeneracy thus scales only as $\exp(c L)$, where $c$ is a constant and $L$ is the linear size of the system.
Since many of the layered states are not only nearly degenerate, but are also separated by a large energy barrier,
the system is stuck in one of the layered states below the charge-ordering transition.

The lack of parametric dependence and the realistic energy hierarchy in our model make it realizable experimentally as  the first mixed coordination ASI and the first extensively degenerate ASI in a critical phase. This phase can  now be accessed  with second generation annealing methods~\cite{morgan11,kapaklis12,nisoli12,greaves12, sheng13, nisoli13}, thus opening novel directions in the study  of frustration-induced degeneracy, dynamics of defects in a critical phase, and because of mixed coordination, monopole excitations in a monopole background~\cite{chern13}.

We thank P.~Mellado, A. Libal, C.~Reichhardt, and R. Moessner for useful comments.
This work was carried out under the auspices of the National Nuclear Security Administration of the U.S. Department of
Energy at Los Alamos National Laboratory under Contract No.~DEAC52-06NA25396. GWC acknowledges the support of LANL Oppenheimer Fellowship.

\section{Supplementary Materials}

\subsection{Smectic phase, charge and spin orders}

\begin{figure*}[t]
\includegraphics[width=1.5\columnwidth]{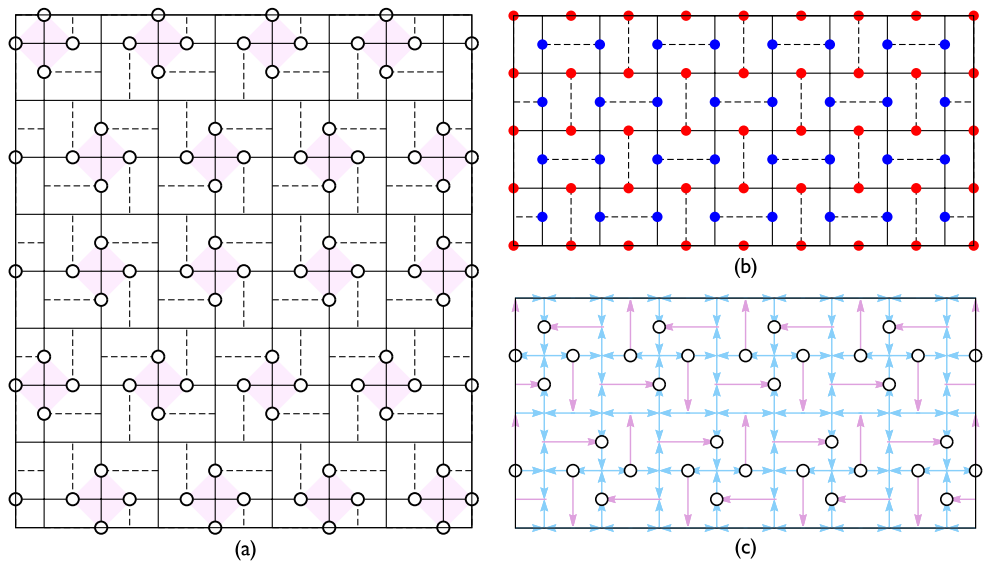}
\caption{(Color online) (a) Quasi-1D ordering of defect vertices in the charge-ordered phase. 
In this case, clusters consisting of four defect-vertices are ordered along the $x$-direction. The 1D ordering has a period of two unit cells.
Depending on the offsets, 1D ordering on different chains can be distinguished by an Ising variable $\tau$.
The configuration shown in panel (a) corresponds to $\tau_y = {1, -1, 1, 1, -1}$. States described by different Ising variables $\{\tau\}$
are exactly degenerate when we consider only the Coulomb interaction between magnetic charges at the vertices. This is
because all such layered states have the same staggered order of magnetic charges shown in Panel (b). Here the blue and red circles
denote vertices with $+1$ and $-1$ magnetic charges (in natural units), respectively.
Because of the sliding symmetry associated with these states, the charge-ordered phase is similar to the well known smectic phase of liquid crystals. 
Inclusion of full dipolar interaction lifts the degeneracy of the layered states. The minimum-energy state shown in Panel (c) corresponds to
a staggered Ising pattern $\{\tau\} = \{\cdots, +1, -1, +1, -1, \cdots\}$.
\label{fig:smectic2}}
\end{figure*}

Here we discuss details of the long-range ordering induced by either the Coulomb interaction
between magnetic charges or the dipolar interaction. As mentioned in the main text, the effective degrees of freedom
in the manifold of effective $F$-model are non-local loops. Single-spin update algorithm
suffers freezing problems at low temperatures. In order to find the long-range ordering at low $T$, we
again employ the loop-algorithm~\cite{evertz93,syljuasen04} discussed in the main text.
With the inclusion of Coulomb or dipolar interactions, however, different loop updates result in either an energy gain or cost.
The energy change of a loop update is computed by first reconstructing the charge or spin configurations
from the effective 6-vertex state, and then calculating the resultant Coulomb or dipolar energy.
A conventional Metropolis criterion is then used to determine whether a loop update is accepted. 

With the aid of the modified loop algorithm, we find that the system undergoes a phase transition 
into a {\em smectic}-like phase at low temperatures. This novel phase [Fig.~\ref{fig:smectic2}(a) and (b)] is characterized by a staggered ordering
of magnetic charges and an intermediate sliding symmetry~\cite{nussinov}.
The charge-order is described by a $Z_2$ Ising order parameter similar to that in kagome spin ice~\cite{moller09,chern11}.
Noting that different spin or defect configurations give rise to the same charge pattern, 
there exists an interesting degeneracy, related to the above sliding symmetries, in the charge-ordered states.
To see this sliding symmetry, we first note that the defect-vertices in this phase `crystalize' along in one particular direction (either $x$ or $y$); see Fig.~\ref{fig:smectic2}(a).
More specifically, the ordering can be viewed as a 1D periodic pattern of 4-defects clusters. Since the ordering has a period of two unit cells,
shifting the whole 1D configuration uniformly by one unit cell gives rise to two distinct 1D patterns which can be distinguished  
by an Ising variable $\tau = \pm 1$.    Consequently, the 2D defect as well as the underlying spin configurations can be
described by a set of Ising variables $\{\cdots, \tau_{n-2}, \tau_{n-1}, \tau_n, \tau_{n+1}, \cdots \}$.
If we consider only Coulomb interactions between magnetic charges at the vertices, 
all such {\em layered} states are energetically degenerate since they have exactly the same staggered charge order shown in Fig.~\ref{fig:smectic2}(b);
the degeneracy is thus $W = 2^L$, where $L$ is the linear size of the system.   It is important to note that this degeneracy is {\em not} extensive. 
It is also worth pointing out that this smectic phase breaks the $C_2$ lattice rotational symmetry.

The degeneracy of the charge-ordered states $\{\tau_n\}$ is lifted by the full dipolar interaction.
The ground state corresponds to a staggering of the Ising variables, i.e. $\{\tau_n\} = \{\cdots, +1, -1, +1, -1, \cdots\}$. The resultant
defect-configuration and spin order is shown in Fig.~\ref{fig:smectic2}(c).
However, we also find that many layered states in the smectic phase are almost degenerate energetically.
More importantly, states described by different Ising variables $\{\tau_n\}$ are separated by huge energy barriers.
In the thermodynamic limit, the system will be stuck in one particular layered state (not necessarily the lowest-energy one shown in Fig.~\ref{fig:smectic2}(c))
as temperature is lowered.

\subsection{Vertex-model energetics}

The existence of the ice phase predicted in the main text depends on an assumption regarding the hierarchy of vertex energies. Intuitively, the hierarchy we use should be the simplest to realize experimentally. Here we present results from micromagnetic simulations which show that this hierarchy can easily be realized using nanoislands much like those in prior experiments (e.g.,~\cite{wang06,morgan11}. Table~\ref{tab:oommf} shows energies for the seven vertex types present in the shakti lattice, computed using the OOMMF micromagnetics package~\cite{oommf}. As expected, \(\epsilon_I<\epsilon_{II}<\epsilon_{III}\) and \(\epsilon_{I^\prime}<\epsilon_{II^{\prime}}<\epsilon_{III^{\prime}}\). Also important is the fact that \(\epsilon_{II}-\epsilon_{I} > \epsilon_{II^{\prime}}-\epsilon_{I^{\prime}}\), meaning Type II\(^\prime\) vertices will be energetically preferable instead of Type II, upon which our entire vertex-model treatment depends.

\begin{table*}
  \begin{tabular}{ l | c c c c c c c }
	Vertex Type & Type I & Type II & Type III & Type IV & Type I\(^\prime\) & Type II\(^\prime\) & Type III\(^\prime\) \\
   Energy (\(10^{-18}\)~J) & -10.6 & -6.02 & -1.48 & 18.1 & -6.70 & -2.47 & 9.12 \\
  \end{tabular}
  \caption{Vertex energies as calculated by the OOMMF package for the vertices shown in Fig.~\ref{Fig:oommf}.}
  \label{tab:oommf}
\end{table*}

\begin{figure}[t]
\includegraphics[width=0.95\columnwidth]{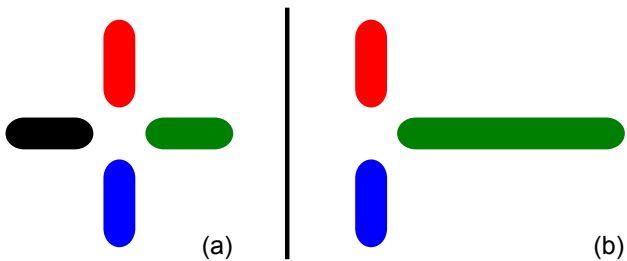}
\caption{(Color online) (a) The mask used for Type I-IV vertices. Each island is 300~nm long by 100~nm wide by 30~nm thick. (b) The mask for Type I\(^\prime\)-III\(^\prime\) vertices. The small islands are as in (a) while the long island is 800~nm in length with other dimensions the same.}
\label{Fig:oommf}
\end{figure}


\begin{thebibliography}{99}

\bibitem{nisoli13} C. Nisoli, R. Moessner, P. Schiffer,  {\it Rev. of Mod. Phys.}, in press.

\bibitem{wang06} R.~F.~Wang, C.~Nisoli, R.~S.~Freitas, J.~Li, W.~McConville, B.~J.~Cooley, M.~S.~Lund, N.~Samarth, C.~Leighton,
V.~H.~Crespi, and P.~Schiffer, Nature {\bf 439}, 303 (2006).

\bibitem{bramwell06} S. T. Bramwell, Nature, {\bf 439}, 273 (2006).

\bibitem{nisoli07} C.~Nisoli, R.~Wang, J.~Li, W.~F.~McConville, F.~E.~Lammert, P.~Schiffer, and V.~H.~Crespi, 
Phys. Rev. Lett. {\bf 98}, 217203 (2007).

\bibitem{nisoli10} C.~Nisoli, J.~Li, X.~Ke, D.~Garand, P.~Schiffer, and V.~H.~Crespi, Phys. Rev. Lett. {\bf 105}, 047205 (2010).

\bibitem{morgan11} J.~P.~Morgan, A.~Stein, S.~Langridge, and C.~H.~Marrows, Nature Phys. {\bf 7}, 75 (2011).

\bibitem{morgan13} J.~P.~Morgan, J.~Akerman, A.~Stein, C.~Phatak, R.~M.~L.~Evans, S.~Langridge, and C.~H.~Marrows,
Phys. Rev. B {\bf 87}, 024405 (2013).

\bibitem{moller09} G.~M\"oller and R.~Moessner, Phys. Rev. B {\bf 80}, 140409 (2009).

\bibitem{chern11} G.-W.~Chern, P.~Mellado, and O.~Tchernyshyov, Phys. Rev. Lett. {\bf 106}, 207202 (2011).

\bibitem{lammert10} P.E.~Lammert, X.~Ke, J.~Li, C.~Nisoli, D.~M.~Garand, V.~H.~Crespi, and P.~Schiffer,
Nature Phys. {\bf 6}, 786 (2010).

\bibitem{ke08} X.~Ke, J.~Li, C.~Nisoli, P.~E.~Lammert, W.~McConville, R.~F.~Wang, V.~H.~Crespi, and P.~Schiffer,
Phys. Rev. Lett. {\bf 101}, 037205 (2008).

\bibitem{mengotti09} E.~Mengotti, L.~J. Heyderman, A.~Bisig, A.~Fraile Rodriguez, L.~Le Guyader, F.~Nolting, 
and H.~B.~Braun, J. Appl. Phys. {\bf 105}, 113113 (2009).

\bibitem{libal06} A.~Lib\'al, C.~Reichhardt, and C.~J.~Olson~Reichhardt, Phys. Rev. Lett. {\bf 97}, 228302 (2006).

\bibitem{tanaka06} M.~Tanaka, E.~Saitoh, H.~Miyajima, T.~Yamaoka, and Y.~Iye, Phys. Rev. B {\bf 73}, 052411 (2006).

\bibitem{qi08} Y.~Qi, T.~Brintlinger, and J.~Cumings, Phys. Rev. B {\bf 77}, 094418 (2008).

\bibitem{li10} J. Li, X. Ke, S. Zhang, D. Garand, C. Nisoli, P. Lammert, V. H. Crespi, and P. Schiffer, Phys. Rev. B {\bf 81}, 092406 (2010).

\bibitem{morrison12} M.~J. Morrison, T.~R. Nelson,  C.~Nisoli, New J. Phys. {\bf 15}, 045009 (2013).

\bibitem{kapaklis12} V. Kapaklis, U. B. Arnalds, A. Harman-Clarke, E. Th.
Papaioannou, M. Karimipour, P.Korelis, A. Taroni P.~C.~W.~Holdsworth, S. T. Bramwell, and B. Hj�oorvarsson, 
New J. Phys. {\bf 14}, 035009 (2012).

\bibitem{nisoli12} C.~Nisoli,  New J. Phys. {\bf  14} 035017 (2012)

\bibitem{chern13} G.-W. Chern and P. Mellado, arXiv:1306.6154 (2013).

\bibitem{greaves12} S.~J.~Greaves and H.~Muraoka, J. Appl. Phys. {\bf 112}, 043909 (2012)

\bibitem{farhan13} A. Farhan,	 P. M. Derlet,	 A. Kleibert,	 A. Balan,	 R. V. Chopdekar,	 M. Wyss,	 L. Anghinolfi,	 F. Nolting	 and L. J. Heyderman Nature Phys. 9, 375Ð382 (2013) 

\bibitem{harris91} M.~J.~Harris, S.~T.~Bramwell, D.~F.~McMorrow, T.~Zeiske, and K.~W.~Godfrey, Phys. Rev. Lett. {\bf 79}, 2554 (1997).

\bibitem{bramwell01} S.~T.~Bramwell and M.~J.~P.~Gingras, Science {\bf 294}, 1495 (2001).

\bibitem{petrenko} V.~F.~Petrenko and R.~W.~Whitworth, {\em Physics of ice} (Oxford University Press, 1999).

\bibitem{bernal33} J. D. Bernal, R. H.  Fowler, J. Chem. Phys. {\bf 1}, 515 (1933). 

\bibitem{pauling35} L. Pauling, J. Am. Chem. Soc. {\bf 57},  2680 (1935).

\bibitem{henley10} C. L. Henley, Annu. Rev. Condens. Matter Phys. {\bf 1}, 179 (2010).

\bibitem{castelnovo08} C.~Castelnovo, R.~Moessner, and S.~L.~Sondhi, Nature {\bf 451}, 42 (2008).

\bibitem{baxter} R.~J.~Baxter, {\em Exactly Solved Models in Statistical Physics} (Academic, New York, 1982).

\bibitem{lieb67} E.~H.~Lieb, Phys. Rev. Lett. {\bf 18}, 1046 (1967);
E.~H.~Lieb and F.~W.~Wu, {\em Phase Transitions and Critical Phenomena} (Academic, London, 1971), Vol.~1.

\bibitem{moller06} G.~M\"oller and R.~Moessner, Phys. Rev. Lett. {\bf 96}, 237202 (2006).

\bibitem{disclaimer} {Inclusion of long range interactions reveals an algebraic ice phase with ordered magnetic charge~\cite{chern11,moller09}. However this phase has not been found experimentally yet, and experimental results point to a satisfying approximation of honeycomb ice in terms of a nearest neighbor vertex model ~\cite{wang06,nisoli07,nisoli10,tanaka06,qi08,li10}.}

\bibitem{cairo} E. Ressouche, V. Simonet, B. Canals, M. Gospodinov, and V. Skumryev, Phys. Rev. Lett. {\bf 103}, 267204 (2009).

\bibitem{parameters} Theoretical modeling of the nanomagnetic arrays is based on either the point-dipole model or 
the dumbbell representation.
Both choices  in this case give the the same energy hierarchy to the vertices of Fig.~\ref{fig:vertex}. Different specific values for $\alpha'$ can be obtained via proper engineering of the middle island. 






\bibitem{evertz93} H.~G.~Evertz, G.~Lana, and M.~Marcu, Phys. Rev. Lett. {\bf 70}, 875 (1993).

\bibitem{syljuasen04} O.~F.~Syljuasen and M.~B.~Zvonarev, Phys. Rev. E {\bf 70}, 016118 (2004).


\bibitem{jaubert11} L.~D.~C.~Jaubert, M.~Haque, and R.~Moessner, Phys. Rev. Lett. {\bf 107}, 177202 (2011).

\bibitem{nussinov} Z. Nussinov, E. Fradkin, Phys. Rev. B {\bf 71}, 195120 (2005); 
Z. Nussinov, C. D. Batista, and E. Fradkin, Int. J. Mod. Phys. B {\bf 20}, 5239 (2006).

\bibitem{supple} See supplementary material for details of the smectic phase.

\bibitem{sheng13} S. Zhang, I. Gilbert, C. Nisoli, GW Chern, M. J. Erickson, L. OÕBrien, C. Leighton, P. E. Lammert1, V. H. Crespi and P. Schiffer {\it Nature} {\bf 500} 553  (2013).  

\bibitem{oommf} The Object Oriented MicroMagnetic Framework (OOMMF) project at ITL/NIST. http://math.nist.gov/oommf/ (2013).


\end{thebibliography}
\end{document}